\documentclass[a4paper, 12pt]{article}

\usepackage[a4paper,top=2cm,bottom=2cm,left=2cm,right=2cm,marginparwidth=1.75cm]{geometry}
\usepackage{amsmath, amssymb, amsfonts, amstext}
\usepackage{graphicx}
\usepackage{natbib}
\usepackage[colorlinks=true, allcolors=blue]{hyperref}
\usepackage{bm}
\usepackage{xcolor}
\usepackage{subcaption}
\usepackage{graphicx}
\usepackage{authblk}
\usepackage{mathptmx}
\usepackage{multirow}

\title{\textbf{\LARGE Cosmological perturbations with $f(R)$ gravity scalarons : Galaxy power spectra and the scalaron mass}}

\author{\small
	Abhijit Talukdar$^{1}$\thanks{abhijittalukdar@gauhati.ac.in},
	Sanjeev Kalita$^{1}$\thanks{sanjeev@gauhati.ac.in},
	\& Shadab Alam$^{2}$\thanks{shadab.alam@tifr.res.in}
	\\
	\small $^{1}$Department of Physics, Gauhati University, Jalukbari, 781014, India\\
    \small $^{2}$Tata Institute of Fundamental Research, Homi Bhabha Road, Mumbai 400005, India
}

\date{}

\begin{document}
\maketitle

\begin{abstract}
In this paper we study cosmological perturbations with $f(R)$ gravity scalaron. We consider {epoch of transition} from early general relativity (GR) phase to late time non-GR phase of cosmogenic history for a viable $f(R)$ gravity Lagrangian close to the $\Lambda$CDM theory at early cosmic history. We study deviation in matter power spectra from $\Lambda$CDM scenario. Galaxy power spectra multipoles are generated using galaxy bias, velocity dispersion and modified gravity parameters. While monopole and quadrupole matter power in $f(R)$ theory increase towards larger $k$ (small scales), quadrupole power spectrum remains elevated relative to $\Lambda$CDM for smaller $k$ (large scales), upto $k\approx 0.02$. These power spectra provide observables for testing gravity through the current and future galaxy survey missions such as DESI, EUCLID, 4MOST and PFS. We report appropriate GR limit of scalaron mass in galactic and massive black hole scales. Evolution of the scalaron mass and its general relativistic limit for various cosmic structures are reported. Deviation from GR in cosmogenic evolution is attributed to decrease in scalaron mass with cosmic time. 
\end{abstract}

\section{\label{sec1}Introduction}

General Relativity (GR) has undergone successful tests in the scale of the solar system \citep{2014LRR....17....4W}, binary pulsars \citep{1975ApJ...195L..51H}, gravitational wave binaries \citep{2016PhRvL.116v1101A}, stellar orbits around the Galactic Center supermassive black hole \citep{PhysRevLett.118.211101} and shadows cast by the Galactic Center black hole \citep{PhysRevLett.116.031101} and the M87 black hole \citep{PhysRevLett.125.141104}. The simplest model of cosmology prevailing in 1990s was based on the maximally symmetric Friedmann-Lemaitre-Robertson-Walker (FLRW) metric applied to general relativistic field equations of gravity where, the source terms are contributed by radiation, baryonic matter and Cold Dark Matter (CDM). It predicted power law expansion of the cosmic scale factor, abundances of light elements, existence of a thermal relic of hot big-bang and evolution of matter density inhomogeneities leading to formation of cosmic structures \citep{1980lssu.book.....P}, However, it was inevitable to extend this simple model to account for the discovery of accelerated expansion of the universe \citep{1998AJ....116.1009R, 1999ApJ...517..565P}. The cosmic speedup has been interpreted as caused by a mysterious dark energy fluid violating the {strong energy conditions}, $\rho+3p \ge 0$. The current standard model of cosmology including cosmic expansion history and evolution of primordial density fluctuations gives meaningful outcomes provided one assumes correctness of GR in cosmological scales and adds a {cosmological constant} $(\Lambda)$ to the CDM component. The $\Lambda$CDM cosmology is, however, plagued with the {cosmological constant problem} \citep{RevModPhys.61.1}, {coincidence problem} \citep{2000astro.ph..5265W} and non detection of Weakly Interacting Massive Particles (WIMPs) which are believed to be sole candidates for CDM \citep{2019JPhG...46j3003S}. The simple extensions of the $\Lambda$CDM cosmology within the framework of GR are the {quintessence} dark energy scenarios \citep{1988ApJ...325L..17P, 1988NuPhB.302..645W, PhysRevLett.80.1582} which involve dynamical scalar field as the dark energy fluid. These scenarios lack strong theoretical foundation as the scalar mass is unnaturally small $(m \sim H_0, \ H_0$ being the Hubble parameter) relative to other scalar fields in particle physics theories \citep{RevModPhys.82.451}. The consequence is that we are yet to be sure on how a gravitation theory works in cosmological scales. This gives sufficient theoretical reason to investigate gravitational theories beyond GR. 

In the context of accelerated expansion, we are concerned with alternative gravity theory which is based on correction to Einstein-Hilbert action in such a way that the Ricci scalar $R$ is replaced by a non-linear function called $f(R)$. These are called $f(R)$ gravity theories. The $f(R)$ function acts as the Lagrangian of the gravitational field. This change in the gravitational action can explain primordial inflation \citep{1980PhLB...91...99S}, late time cosmic acceleration \citep{PhysRevD.70.043528, RevModPhys.82.451, PhysRevD.68.123512, 2002IJMPD..11..483C} and flat rotation curves of galaxies \citep{2007MNRAS.375.1423C}. It has been found that $f(R)\approx R^n (n>0)$ leads to modification of primordial inflation and $f(R)\approx R^{-m} (m>0)$ gives rise to late time de-Sitter expansion. In \cite{2003PhLB..576....5N}, it is demonstrated that the time dependent compactification of higher dimensional M/string theory can give rise to correction to GR that looks like the latter class of $f(R)$ theories. Theories with $f(R)$ function as the gravitational Lagrangian are simple to compare with GR based field equations. In addition, these theories are unique among other higher order gravitational theories with the Lagrangian $L=f(R, R_{\mu\nu}R^{\mu\nu}, C_{\mu \nu \alpha \beta}C^{\mu \nu \alpha \beta})$ (where, $C$ is the Weyl tensor) in the sense that they are free from Ostrogradsky instability \citep{Woodard2007}. {Recent DESI DR2 Baryon Acoustic Oscillation (BAO) studies have shown strong statistical evidence in the favor of $f(R)$ models over the standard $\Lambda$CDM scenario \citep{gtrg-56fj}}.

In contrast to GR, $f(R)$ gravity theory contains an additional spin-0 gravitational degree of freedom known as scalaron. It is defined by the scalar field $\psi=f'(R)$. Success of GR in the solar system naturally calls for the {chameleon mechanism} \citep{2020PhRvD.101f4005N} which increases the scalaron mass in high density environments such as solar and planetary densities \citep{Amendola_Tsujikawa_2010}. In earlier studies spherically symmetric \citep{2018ApJ...855...70K, 2020ApJ...893...31K} and stationary \citep{2024ApJ...964..127P}, axisymmetric vacuum solutions of $f(R)$ gravity field equations \citep{2018ApJ...855...70K, 2024ApJ...964..127P} were developed. Influence of scalarons on periapsis shift of stellar orbits near the Galactic Center black hole \citep{2023IJMPD..3250021P}, black hole shadow size \citep{2024ApJ...964..127P}, Big-Bang Nucleosynthesis \citep{2024JCAP...02..019T, 2024ApJ...970...91T} and planetary orbits \citep{2025PhyS..100f5006P} have been studied. A converging GR limit of scalaron mass has been obtained as $m_\psi \ge 10^{-16}$ eV. Scalaron presents a Yukawa correction to the Newtonian gravitational potential \citep{2018ApJ...855...70K}, thereby, producing an extra additive gravitational force.  

Given the success of the $\Lambda$CDM cosmology in explaining cosmological observations \citep{2020A&A...641A...6P}, we believe that a cosmological alternative to GR must look like $\Lambda$CDM. Cosmogenic impact of scalaron can be interesting as it affects growth of {Large Scale Structures} (LSS) \citep{Amendola_Tsujikawa_2010}. The growth of LSS is sensitive to the behavior of gravitational force. The evolution of matter density fluctuation ($\delta \rho_m$) and hence, the curvature $(\Phi)$ and the Newtonian potential $(\Psi)$ depend on the Poisson equation which gets modified in an alternative gravitation theory. Perturbation of the scalaron field $(\delta \psi)$ and the scalaron mass $(m_\psi)$ together affect the growth of LSS. The scalaron mass naturally depends on the Ricci scalar and hence the cosmological background density. The universe has successfully crossed a general relativistic and high density matter dominated phase after CMB decoupling. Therefore, the dominance of Newtonian potential over the Yukawa correction demands that the scalaron must be heavy in the high redshift (early) universe. In the cosmological perturbation theory, we compare the scalaron mass with the Fourier mode of the fluctuation, $k/a$, where $a$ is the cosmic scale factor. {Heavy and light scalarons are characterized by $m_\psi>>k/a$ and $m_\psi<<k/a$ respectively}. Therefore, transition from GR phase of LSS history to its non-GR phase is given by the condition $m_\psi(a)=k/a$ \citep{Amendola_Tsujikawa_2010}.

In this paper, we consider one $f(R)$ gravity Lagrangian which reproduces the standard $\Lambda$CDM cosmology at high redshift (deep matter era) and calculate scalaron mass at the transition epoch. We also find that {GR phase terminates quite early in cosmic history} ($z=5-15$) for supermassive black hole scales of cosmological fluctuations. We show that this phase can be longer for gravity theory constrained by observed upper limit of deviation of the spectral index $(n_s)$ in the matter power spectra, $P_k \sim k^{n_s}$. Evolution of scalaron mass from early to late universe is also presented.

Section \ref{sec2} presents the role of scalarons in cosmological perturbation. In section \ref{sec3} we show deviation from $\Lambda$CDM power spectrum with the help of of monopole and quadrupole power of galaxies and present scalaron mass from for different cosmic structures and its evolution. Section \ref{sec4} contains discussion of main results.

\section{\label{sec2} $f(R)$ gravity Lagrangian and cosmological perturbations with scalarons}

\subsection{Choice of the gravitational Lagrangian}

The $f(R)$ gravitational theories are expressed by the modified Einstein-Hilbert action (in the unit $8\pi G =1$ and $c=1$),
\begin{equation}
	S=\frac{1}{2}\int d^4x \sqrt{-g}f(R) + \int \mathcal{L}_m d^4x \sqrt{-g}, 
\end{equation}
where, $\mathcal{L}_m$ is the Lagrangian of the non-gravitational source. In the metric formalism, the variation of the above action with respect to $g_{\mu\nu}$ gives the following field equation,
\begin{equation}\label{eq2}
	\psi R_{\mu\nu}-\frac{1}{2}f(R) g_{\mu\nu}= T_{\mu\nu}+\nabla_\mu \nabla_\nu \psi-g_{\mu\nu}\Box{\psi}.
\end{equation}

Correction to GR is written as,

\begin{equation}
	f(R)=R-2\Lambda+g(R),
\end{equation}
where, $\Lambda$ is the cosmological constant and $g(R)$ is the correction term. In \cite{2003PhLB..576....5N}, it was shown that these curvature corrections can originate from time dependent compactification of $4+n$ dimensional gravity in string theory. The starting proposals which invoked new gravitational physics as the cause of late time cosmic acceleration contain correction terms of type $g(R)\varpropto R^{-m} (m>0)$ \citep{PhysRevD.70.043528, 2003PhLB..576....5N}. These models suffer from instabilities associated with $f''(R)<0$ \citep{2003PhLB..575....1C, 2003PhLB..573....1D} and violate solar system tests of GR \citep{2003PhLB..573....1D, PhysRevD.72.083505}. Here, prime ($'$) denotes derivative with respect to Ricci scalar $R$. We consider two gravitational Lagrangians which satisfy local gravity tests \citep{PhysRevD.68.023522, 2004MPLA...19.1509S}, presence of matter dominated era before the onset of cosmic accelerated expansion \citep{PhysRevLett.98.131302, PhysRevD.75.083504} and stability of cosmological perturbations \citep{2006NJPh....8..323C, PhysRevD.75.044004}. The Lagrangians are given by \citep{Amendola_Tsujikawa_2010},
\begin{align}
	f(R)&=R-\mu R_c\frac{(R/R_c)^{2n}}{(R/R_c)^{2n}+1},\\
	f(R)&=R-\mu R_c \left[1-(1+R^2/R_c^2)^{-n}\right].
\end{align}

Here, $\mu, \ R_c, \  n>0$. $R_c$ is a characteristic scalar curvature so that in the high density phase of the early universe ($z>>1$), $R>>R_c$ and both the Lagrangians reduce to GR like Lagrangian,
\begin{equation}
	f(R) \approx R-\mu R_c \ .
\end{equation}

We identify $\mu R_c$ as $2\Lambda$. We adopt the following common form of the two Lagrangians for the high $z$ universe,
\begin{equation}\label{eq7}
	f(R) \approx R-\mu R_c \left[1-\left(\frac{R}{R_c}\right)^{-2n}\right].
\end{equation}

The above Lagrangian shows that the correction to general relativistic $\Lambda$CDM cosmology occurs in the low curvature regime, $R\le R_c$.

The mass of the scalaron is defined as,
\begin{equation}
	m_\psi^2=\frac{f'(R)}{3 f''(R)}.
\end{equation}

Throughout our calculations we take $f'(R)\approx 1$ as the viable $f(R)$ models have to be very close to the general relativistic $\Lambda$CDM cosmology.

\subsection{Evolution of perturbations with scalarons}

The perturbed flat space FLRW metric is given as \citep{Amendola_Tsujikawa_2010},
\begin{equation}
	ds^2=e^{2N}[-(1+2\Psi)\mathcal{H}^{-2}dN^2+(1+2\Phi)\delta_{ij}dx^i dx^j],
\end{equation}
where, $\mathcal{H}=aH$. $\Psi$ and $\Phi$ are Newtonian potential and curvature potential respectively and $N=\ln a$.

Modification to GR produces dark anisotropic stress, $\Phi_+=1/2(\Phi+\Psi)$, which is expressed as (with $f'(R)=1$),
\begin{equation}
	\Phi_+ \varpropto f''(R).
\end{equation}

Deviation from GR can be quantified by the deviation parameter,
\begin{equation}\label{eq11}
	B \varpropto\frac{dR}{d \ln H}\Phi_+ .
\end{equation}

The Integrated Sachs–Wolfe (ISW) potential is given by,
\begin{equation}\label{eq12}
	\Phi_-=\frac{1}{2}(\Phi-\Psi).
\end{equation}

It measures how much the weak lensing signal deviates from GR like potential $\Phi=-\Psi$.

We consider the presence of non-relativistic matter in the metric $f(R)$ gravity formalism. Density perturbation equations for pressureless matter ($\omega_m=0=c_s^2$, $c_s$ being the sound speed) are given by \citep{Amendola_Tsujikawa_2010},
\begin{align}
	\delta_m'&=-(\theta_m+3\Phi') \label{eq13}, \\
	\theta_m'&=\Psi/\hat{\lambda}^2-(1+\mathcal{H}'/\mathcal{H})\theta_m \label{eq14},
\end{align}
where, $\delta_m=\delta \rho/\rho$ and $\theta_m=i \vec{k}.\vec{v}$ are the matter density contrast and velocity divergences respectively. Here prime ($'$) denotes derivative with respect to $\ln a$. $\hat{\lambda}=\mathcal{H}/k=aH/k$. Differentiating equation \eqref{eq13} with respect to $\ln a$ and using equation \eqref{eq14}, we get, 
\begin{equation}
	\delta_m''+\left(1+\frac{\mathcal{H}'}{\mathcal{H}}\right)\delta_m'+\frac{1}{\hat{\lambda}^2}\Psi=-3\left(1+\frac{\mathcal{H}'}{\mathcal{H}}\right)\Phi'-3\Phi''.
\end{equation}

For the fluctuation modes deep inside the Hubble radius i.e. $\hat{\lambda}<<1$, the above equation reduces to,
\begin{equation}
	\delta_m''+\left(1+\frac{\mathcal{H}'}{\mathcal{H}}\right)\delta_m'+\frac{1}{\hat{\lambda}^2}\Psi=0.
\end{equation}

By perturbing equation \eqref{eq2}, we get the perturbation in the scalaron field $\delta \psi$. Perturbed equations in the Fourier space are obtained (in the unit of $8\pi G=1$) as \citep{PhysRevD.71.063536},

\begin{equation}\label{eq17}
	-\frac{k^2}{a^2}\Phi+3H(H\Psi-\dot{\Phi})= \frac{1}{2\psi}\left[3H\dot{\delta\psi}-\left(3\dot{H}+3H^2-\frac{k^2}{a^2}\right) \delta \psi-3H\dot{\psi}\Psi-3\dot{\psi}(H\Psi-\dot{\Phi})-\delta \rho_m\right],
\end{equation}

\begin{equation}
	\label{eq18}
	\ddot{\delta \psi}+3H \dot{\delta \psi}+\left(\frac{k^2}{a^2}-\frac{R}{3}\right)\delta \psi = \frac{1}{3}\delta\rho_m+\dot{\psi}(3H\Psi+\dot{\Psi}-3\dot{\Phi})
	+(2\ddot{\psi}+3H\dot{\psi})\Psi-\frac{1}{3}\psi\delta R,
\end{equation}

\begin{equation}\label{eq19}
	\Psi+\Phi=-\frac{\delta\psi}{\psi},
\end{equation}

where, dot $(\dot{})$ represents derivative with respect to the cosmic time. Considering the modes deep inside the Hubble radius $(k^2/a^2>>H^2)$ and taking $k^2/a^2$ and $\delta\rho_m$ as dominant terms, we obtain potentials from equations \eqref{eq17} and \eqref{eq19},
\begin{equation}\label{eq20}
	\Phi=\frac{1}{2\psi}\left(\frac{a^2}{k^2}\delta\rho_m-\delta \psi\right), \  \ \ \  \Psi=-\frac{1}{2\psi}\left(\frac{a^2}{k^2}\delta\rho_m+\delta \psi\right).
\end{equation}

Clearly, $\psi=1$ and $\delta\psi=0$ correspond to GR like situation $\Phi=-\Psi$. In flat FLRW spacetime Ricci scalar can be expressed in terms of Hubble parameter as, $R=6(\dot{H}+2H^2)$. Assuming a power law variation of the scale factor $a(t)\varpropto t^p$, it can be easily shown that $R=6(\dot{H}+2H^2) \sim H^2$. Further assuming extremely slow variation of the scalaron field as compared to the Hubble time i.e. $|\dot{\psi}|\lesssim|H\psi|$ and $|\ddot{\psi}|\lesssim |H^2\psi|$, we neglect all the terms of equation \eqref{eq18} except $\delta\rho_m$ and $(k^2/a^2)\delta\psi$. This is justified as follows. For $\psi=f'(R)$, $\dot{\psi}/\psi=\dot{R}/3m_\psi^2$. As $R\sim H^2$ and $t^{-1}\sim H$, $\dot{\psi}/\psi H\approx (2/3)(H/m_\psi)^2$. In the same spirit, one finds that $\ddot{\psi}=H \dot{\psi}$ leading to $\ddot{\psi}/H^2 \psi=\dot{\psi}/\psi H$. A slow variation of the scalaron field is possible for $m_\psi>>H$, or when the Compton wavelength of the scalaron is well inside the Hubble radius $\sim H^{-1}$. This condition enables us to compare $m_\psi$ with wavenumber $k/a$ of modes deep inside the Hubble radius. The equation for evolution of $\delta \psi$ is, then, written as,
\begin{equation}\label{eq21}
	\ddot{\delta\psi}+3H\dot{\delta\psi}+\left(\frac{k^2}{a^2}+m_\psi^2\right)\delta\psi=\frac{1}{3}\delta\rho_m.
\end{equation}

The $f(R)$ models considered for the study reduce to a common form in the high $z$ universe (see equation \eqref{eq7}). For that specific form of $f(R)$, the time variation of scalaron mass is obtained as,
\begin{equation}\label{eq22}
	m_\psi \simeq \sqrt{\frac{1}{3}\frac{1}{f''(R)}} \simeq t^{-2(n+1)}.
\end{equation}

{The scalaron mass is higher in the past and gradually decreases towards the present epoch. Hence, the equation \eqref{eq21} can be solved for two different cases: a) $m_\psi^2>>k^2/a^2$ and b) $m_\psi^2<<k^2/a^2$.

Now we study the evolution of matter perturbation for the regime of massive scalaron $m_\psi^2>>k^2/a^2$. We start the analysis of perturbations by solving equation \eqref{eq21}. The general solution of the equation can be written as the sum of two solutions namely, the {oscillatory solution} and the {induced solution} as,
\begin{equation}
	\delta\psi=\delta\psi_\text{osc}+\delta\psi_\text{ind}.
\end{equation}

To obtain the oscillatory solution, matter perturbation part is set to zero, $\delta\rho_m=0$. The equation can be rewritten as,
\begin{equation}
	\ddot{\delta\psi_\text{osc}}+3H\dot{\delta\psi_\text{osc}}+m_\psi^2\delta\psi=0.
\end{equation}

WKB approximation can be used express the solution of the above equation and is given as,
\begin{equation}
	\delta\psi_\text{osc} = c_0  a^{-3/2}m_\psi^{-1/2} \cos\left[c_1\int m_\psi(t)dt\right].
\end{equation}

Curvature perturbation varies as, $\delta R \varpropto m_\psi^2 \delta\psi_{\text{osc}}$. Therefore, in deep matter era with $R \sim t^{-2}$, $a\sim t^{2/3}$,
\begin{equation}
	\delta R \approx c_0 t^{-(3n+4)}\cos\left[c_1\int m_\psi(t)dt\right].
\end{equation}

Clearly, $\delta R/R \sim t^{-(3n+2)}$. It diverges to infinity in the past. For stability of perturbations, we have to consider $c_0, \ c_1 \rightarrow 0$. This leaves us with only the induced solution.

Again, the induced solution is through quasi-static approximations where, the variation of the field is assumed to be extremely slow as compared to the source. In this solution, the response of the field to the matter density contrast is considered. Equation \eqref{eq21} reduces to,
\begin{equation}
	m_\psi^2 \delta\psi_\text{ind} \simeq \frac{1}{3}\delta\rho_m.
\end{equation}

For modes deep inside the Hubble radius, this gives,
\begin{align}
	\delta_m \varpropto t^{2/3}, \label{eq28} \\
	\delta \psi_\text{ind} \varpropto t^{4\left(n+\frac{2}{3}\right)}, \\
	\delta R_\text{ind} \simeq \delta\rho_m \varpropto t^{-4/3}.  
\end{align}

It is quite evident that the perturbation $\delta R_\text{ind}/R$ is stable and becomes important in the late time. Also, the matter perturbation growth exactly evolves as the general relativistic scenario. Therefore, in the regime $m_\psi^2>>k^2/a^2$, the matter growth mimics GR like evolution.

As it is seen that the oscillatory mode gets suppressed for $m_\psi^2<<k^2/a^2$. There we consider only the matter induced solution and equation \eqref{eq21} takes the form,
\begin{equation}
	\delta\psi_\text{ind}\simeq \frac{a^2}{3k^2}\delta\rho_m.
\end{equation}

The evolution of matter perturbation and curvature perturbation is given as,
\begin{align}
	\delta_m \varpropto t^{\frac{\sqrt{33}-1}{6}}, \label{eq32} \\
	\delta\psi_\text{ind} \varpropto t^{\frac{\sqrt{33}-5}{6}}, \\
	\delta R_\text{ind} \simeq \delta\rho_m \varpropto t^{-4n+\frac{\sqrt{33}-29}{6}}.
\end{align}

In this regime, the matter perturbation evolution is more significant than in the regime $m_\psi^2>>k^2/a^2$. This indicates that as the universe makes transition from $m_\psi^2>>k^2/a^2$ to $m_\psi^2<<k^2/a^2$, matter experiences enhanced growth,
\begin{equation}
	\frac{\delta_m(m_\psi^2<<k^2/a^2)}{\delta_m(m_\psi^2>>k^2/a^2)} \sim t^{0.124}.
\end{equation}
}

It is worthy of discussing the meaning of GR and non-GR phase in terms of the behavior of gravity. We proceed as follows.

Equation \eqref{eq20} gives us,
\begin{equation}
	\frac{k^2}{a^2} \Psi=-\frac{1}{2}\left[\delta\rho_m+\frac{k^2}{a^2}\delta\psi\right].
\end{equation}

Considering slow varying scalaron field perturbation in equation \eqref{eq21} we get,
\begin{equation}
	\delta\psi=\frac{1}{3}\frac{\delta\rho_m}{m_\psi^2+\frac{k^2}{a^2}}.
\end{equation}

This gives the Poisson equation of gravity as,
\begin{equation}
	\frac{k^2}{a^2} \Psi=-\frac{1}{2}\left[1+\frac{k^2/3a^2}{m_\psi^2+(k^2/a^2)}\right]\rho_m\delta_m.
\end{equation}

We compare the above equation with the conventional Poisson equation to obtain an effective gravitational constant given by (in appropriate unit of $G$), 
\begin{equation}\label{eq_Geff}
	G_\text{eff}=G\left[1+\left(\frac{k^2/3a^2}{m_\psi^2+(k^2/a^2)}\right)\right].
\end{equation}

The low mass scalarons ($m_\psi \rightarrow 0$) gives $G_\text{eff}=(4/3) G$, the maximal deviation from GR. On the other hand, very high mass scalarons ($m_\psi>>k/a$) recovers GR, $G_\text{eff}= G$. The strength of gravity is enhanced when scalaron mass comes down through the transition epoch. This is the onset of modified gravity era.

We define $\nu(a,k)=G_\text{eff}/G$ and this gives the matter perturbation equation as,
\begin{equation}
	\ddot{\delta_m}+2 H \dot{\delta_m}=\frac{3}{2}H^2\nu(a,k)\delta_m.
\end{equation}

During matter dominated era ($a \varpropto t^{2/3}$) we consider a power law ansatz for matter perturbation, $\delta_m \varpropto t^p$. This gives the following constraint to be satisfied by the exponent $p$,
\begin{equation}
	p^2+\frac{p}{3}-\frac{2\nu}{3}=0,
\end{equation}

The growing mode solution is obtained from the root,
\begin{equation}
	p=\frac{-1+\sqrt{1+24\nu}}{6}.
\end{equation}

Growth rate is obtained from,
\begin{equation}
	f^{f(R)}=\left(\frac{d \ln \delta_m}{d \ln t}\right)\left(\frac{d \ln t}{d \ln a}\right).
\end{equation}

Here, $\left({d \ln \delta_m}/{d \ln t}\right)=p$ and $\left({d \ln t}/{d \ln a}\right)=3/2$. This gives the modified growth rate as,
\begin{equation}\label{mod_f}
	f^{f(R)}=\frac{\sqrt{1+24 \nu(a,k)}-1}{4}.
\end{equation}

It is seen that the growth rate in $f(R)$ gravity is a function of scalaron mass ($m_\psi$) and scale ($k$) {through the function $\nu$ (see the defintion of $\nu$ and equation~\eqref{eq_Geff}). Therefore, any observational constraint on $f(R)$ gravity growth rate will be a constraint on the scalaron mass at a given scale $k$.}

\section{\label{sec3} Power spectra and scalaron mass}

In this section we study matter power spectrum in $\Lambda$CDM and $f(R)$ gravity theory with observationally permitted departure from Harrison-Zeldovich spectral index. With a relation between this departure and the parameter $n$ in the $f(R)$ gravity Lagrangian, {we estimate the transition epoch for GR to non-GR phase of evolution of the universe}. We also study the time evolution of ISW potential and evolution of scalaron mass.

{The evolution of scalaron mass with respect to cosmic time (see equation \eqref{eq22}) marks the transition of different structures from $m_\psi^2>>k^2/a^2$ to $m_\psi^2<<k^2/a^2$. We identify these as GR and non-GR phases respectively}. As we have seen, this gives enhanced matter perturbation in the non-GR phase of evolution. The era at which $m_\psi=k/a$, is given by the cosmic time,
\begin{equation}\label{eq40}
	t_k \varpropto k^{\frac{-3}{6n+4}}.
\end{equation}

This relation marks a very significant property of evolution of perturbations. The smaller scale modes enter the non-GR evolution faster than the larger scales.

For the models given by equation \eqref{eq7}, we obtain,
\begin{equation}
	m_\psi^2 \varpropto (\mu R_c)^{-1}[2n(2n+1)]^{-1}\left(\frac{R}{R_c}\right)^{2(n+1)}.
\end{equation}

If $z_k$ is the cosmological redshift corresponding to the transition $m_\psi=k/a$, then the relation $(1+z)=a^{-1}$ gives us that transition redshift as,

\begin{equation}\label{eq38}
	z_k \simeq \left[\left(\frac{k}{a_0H_0}\right)^2 \frac{2n(2n+1)}{\mu^{2n}} \frac{(2\Omega_{DE}^{(0)})^{2n+1}}{\Omega_m^{(0)2(n+1)}}\right]^{\frac{1}{6n+4}} - 1.
\end{equation}

Here, $a_0=1$ is the present scale factor and $\Omega_\text{DE}^{(0)}=\rho_\text{DE}^{(0)}/\rho_\text{crit}^{(0)}$, the present dark energy density parameter. Transition of the universe from a higher mass scalaron to a low mass regime allows us to study interesting scalaron masses at different astrophysical and cosmological scales. Those scalaron masses can be obtained using $m_\psi=k(1+z_k)$.

The value of the parameter $n$ is not yet guided by a fundamental theory. However, one gets clue from constraint on the matter power spectrum of density fluctuations. For this we compute the power spectrum in $f(R)$ gravity theory ($P_{\delta_m}^{f(R)}(k)$) as follows: from equations \eqref{eq28} and \eqref{eq32}, we find that $P_{\delta_m}^{f(R)}(t)$ and $P_{\delta_m}^{\Lambda\text{CDM}}(t)$ are related at any cosmic time $t$ as,
\begin{equation}\label{eqP_diff}
	\frac{P_{\delta_m}^{f(R)}(t)}{P_{\delta_m}^{\Lambda\text{CDM}}(t)}=\left(\frac{t}{t_k}\right)^{\frac{\sqrt{33}-5}{3}}.
\end{equation}

The above relation is normalized in such a way the two power spectra match at the transition epoch $t=t_k$. As $t_k \varpropto (1+z_k)^{-3/2}$, equation \eqref{eqP_diff} shows that,
\begin{equation}\label{eqMPSp}
	P_{\delta_{m}}^{f(R)}(k) \propto k^{n_s+\frac{\sqrt{33}-5}{6n+4}}.
\end{equation}

Here, we have used $P_{\delta_{m}}^{\Lambda \text{CDM}}(k) \sim k^{n_s}$ with $n_s$ being the Harrison-Zeldovich spectral index in $\Lambda$CDM cosmology. Thus, we obtain a difference in the spectral index $n_s$ of the power spectra of both gravity theories,
\begin{equation}
	\Delta n_s = {\frac{\sqrt{33}-5}{6n+4}}.
\end{equation}

\begin{figure}[!ht]
	\centering
	\includegraphics[width=0.7\textwidth]{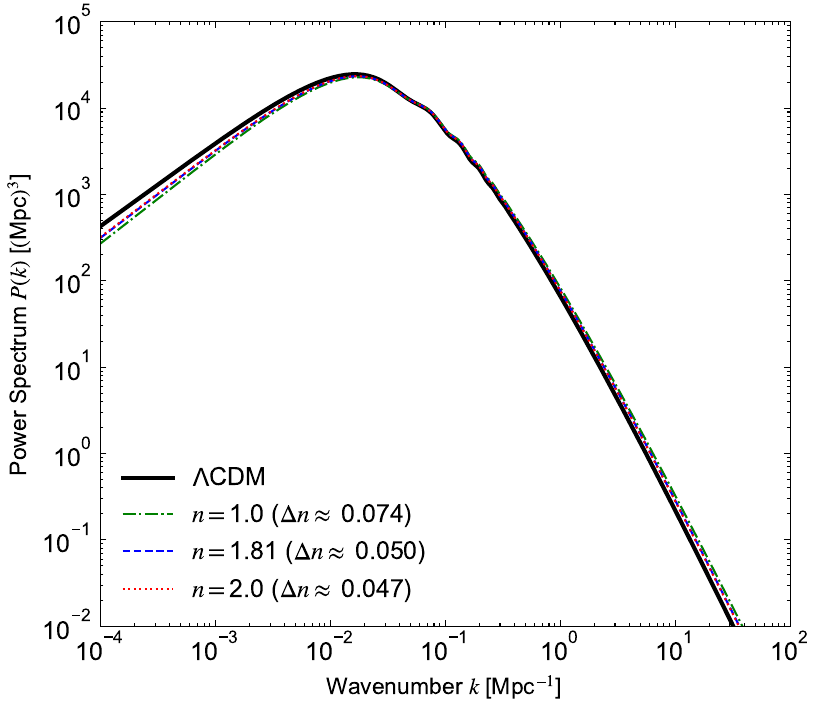}
	\caption{\label{fig1} Theoretical matter power spectra for GR and non-GR scenario (parameterized by $n$). Coloured lines represent power spectra in the $f(R)$ gravity scenario for different values of $n$. Power spectra are computed using \texttt{CAMB} \citep{2000ApJ...538..473L}.}
\end{figure}

$n$ values determine how rapidly scalaron mass changes with cosmic time.
\begin{equation}
	m_\psi \varpropto \rho^{(n+1)}
\end{equation}

This gives,
\begin{equation}
	\frac{\Delta m_\psi}{\Delta \rho} \varpropto (n+1) \rho^n
\end{equation}

At a given epoch (fixed $\rho$), a larger value of $n$ produces steep slope of $m_\psi$ vs $\rho$. Therefore, $n$ determines how fast {GR phase undergoes transition to non-GR phase (higher $m_\psi$ to lower $m_\psi$)}. 

Observationally, there is not much difference in the slope of LSS and CMB. Numerical simulations show little difference in the $\Delta n_s$ values during the present epoch and the epoch of cosmic acceleration. It allows to put constraint on the value of $n$. In order to satisfy local gravity constraint we consider the bound $n>0.9$ \citep{Amendola_Tsujikawa_2010}. Further, LSS+CMB constraint demands $\Delta n_s \le 0.05$, which gives $n\ge 1.81$ \citep{PhysRevD.77.023507}. For our further analysis, we considered a value $n=1$. For the marginal value $\Delta n_s = 0.05$, we take $n=1.81$. We also consider $n=2$. The power spectra for $\Lambda$CDM and $f(R)$ cosmologies are displayed in Figure~\ref{fig1}. Observationally testable power spectra, however, depend on galaxy biases, velocity dispersion and modified gravity parameters. This has been discussed in section~\ref{sec3.1} (see figures~\ref{fig:combined_vertical_1}--\ref{fig:combined_vertical_3}).

\subsection{Galaxy power spectra}\label{sec3.1}

\begin{figure}[!ht]
	\centering
	\begin{subfigure}{\textwidth}
		\centering
		\includegraphics[width=\linewidth]{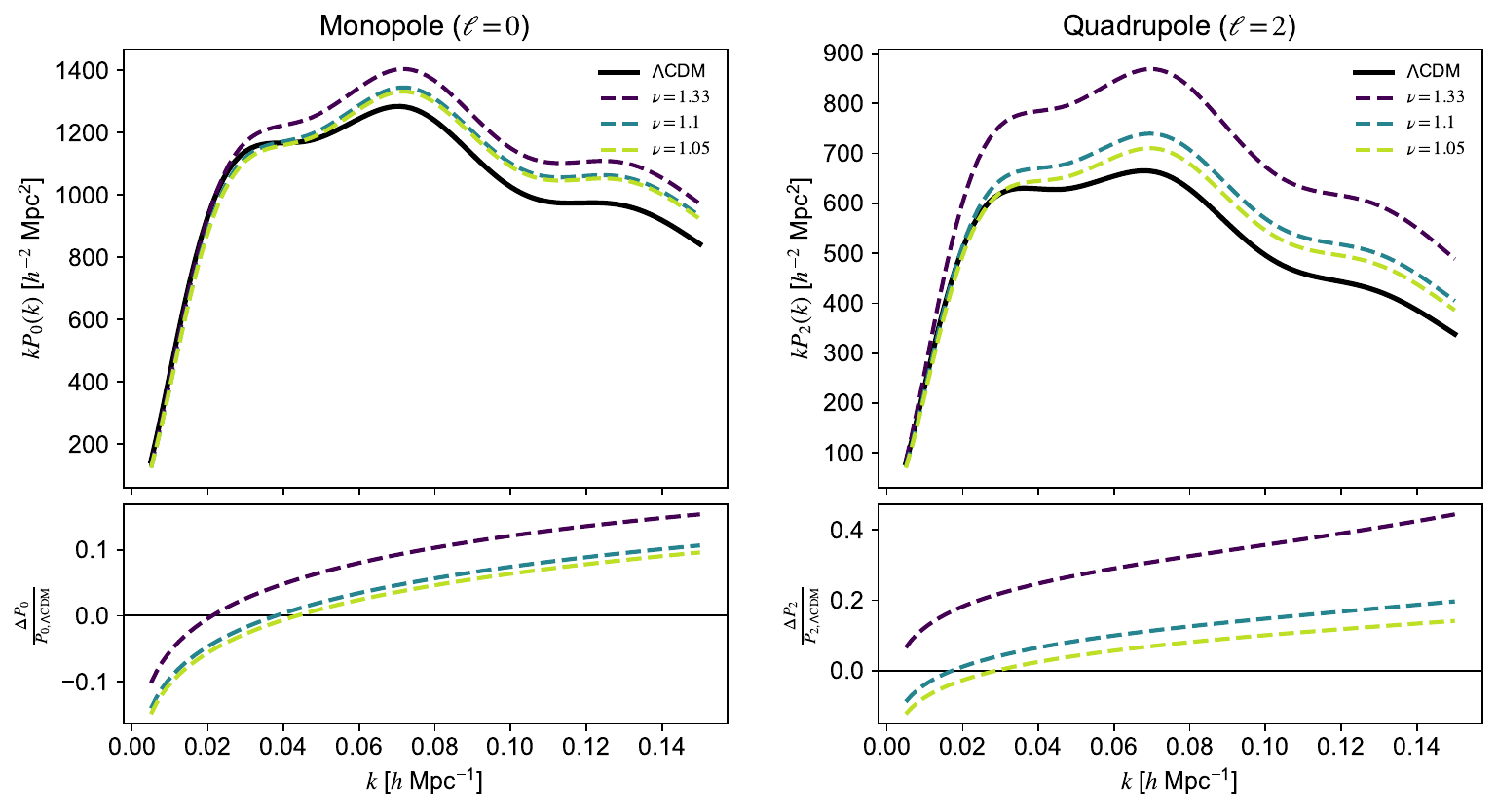}
		\caption{}
		\label{fig_d_1}
	\end{subfigure}

	\begin{subfigure}{\textwidth}
		\centering
		\includegraphics[width=\linewidth]{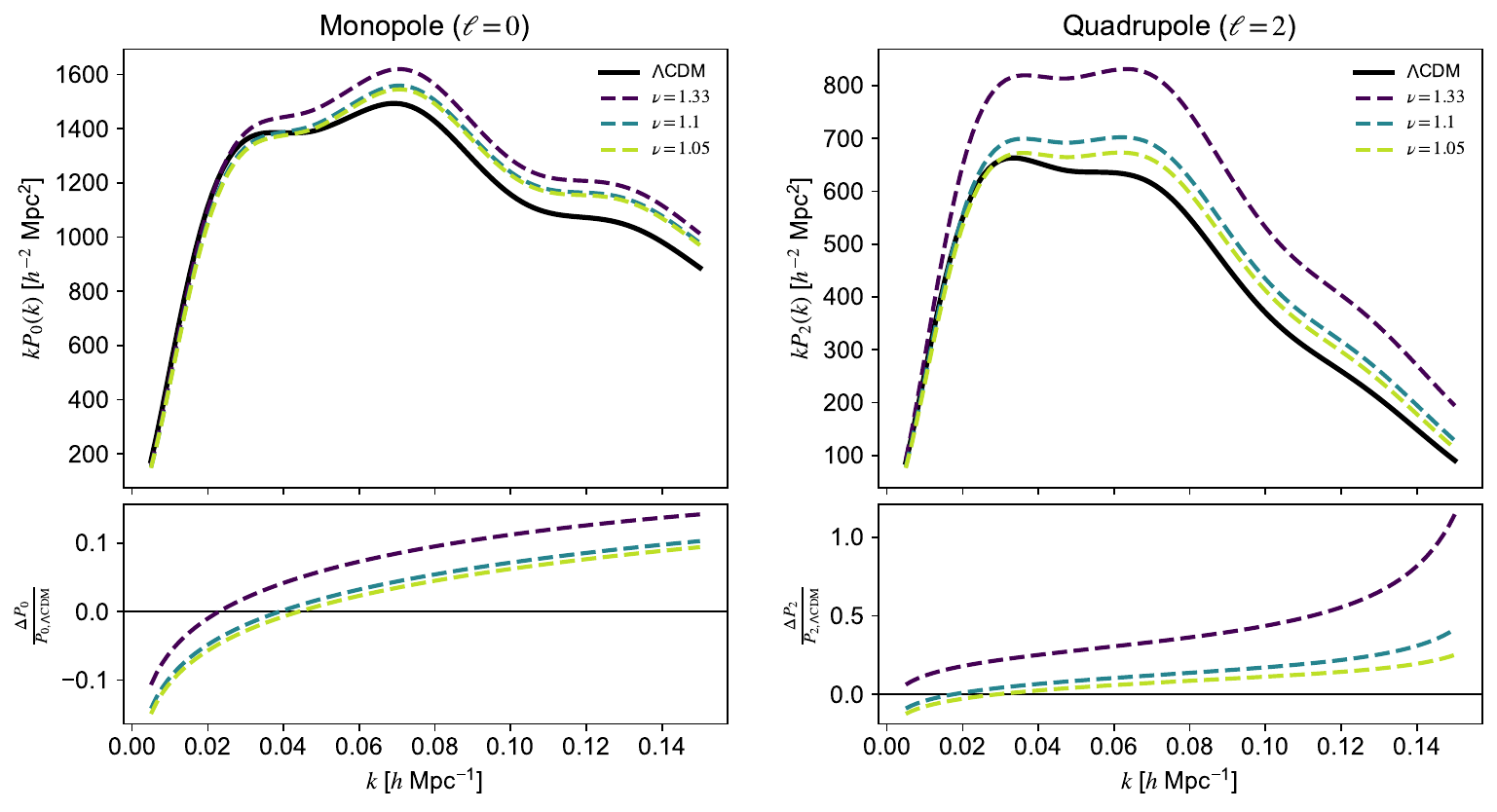}
		\caption{}
		\label{fig_d_2}
	\end{subfigure}

	\caption[Combined analysis of multipoles]{Combined analysis of multipoles. All the power spectra are evaluated at redshift $z=0.8$. Here, $\Delta P_\ell=(k P_{\ell, f(R)}-k P_{\ell, \Lambda\text{CDM}})$. Power spectra multipoles are generated for $n=1$. Panel (a) shows for $b=1.8$ and $\sigma_p=3$ Mpc/h, while Panel (b) illustrates for $b=2$ and $\sigma_p=5$ Mpc/h. Coloured lines represent $\nu = 1.33, \ 1.10$ and $1.05$ respectively and the colours are depicted in the top right panels of each Figure in Figures~\ref{fig_d_1} and~\ref{fig_d_2}.}
	\label{fig:combined_vertical_1}
\end{figure}

\begin{figure}[!ht]
	\centering
	
		\begin{subfigure}{\textwidth}
		\centering
		\includegraphics[width=\linewidth]{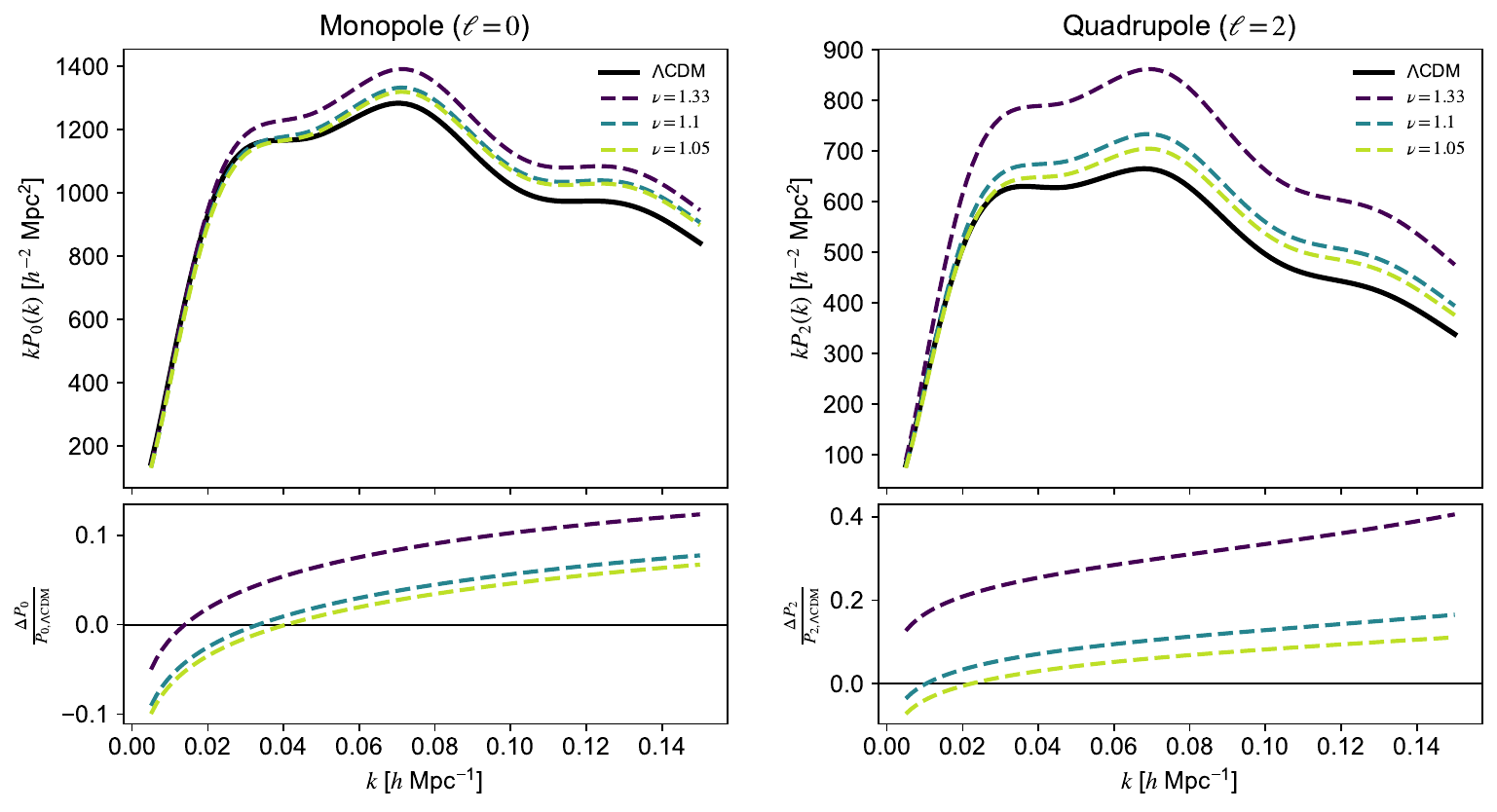}
		\caption{}
		\label{fig_d_3}
	\end{subfigure}
	
	\begin{subfigure}{\textwidth}
		\centering
		\includegraphics[width=\linewidth]{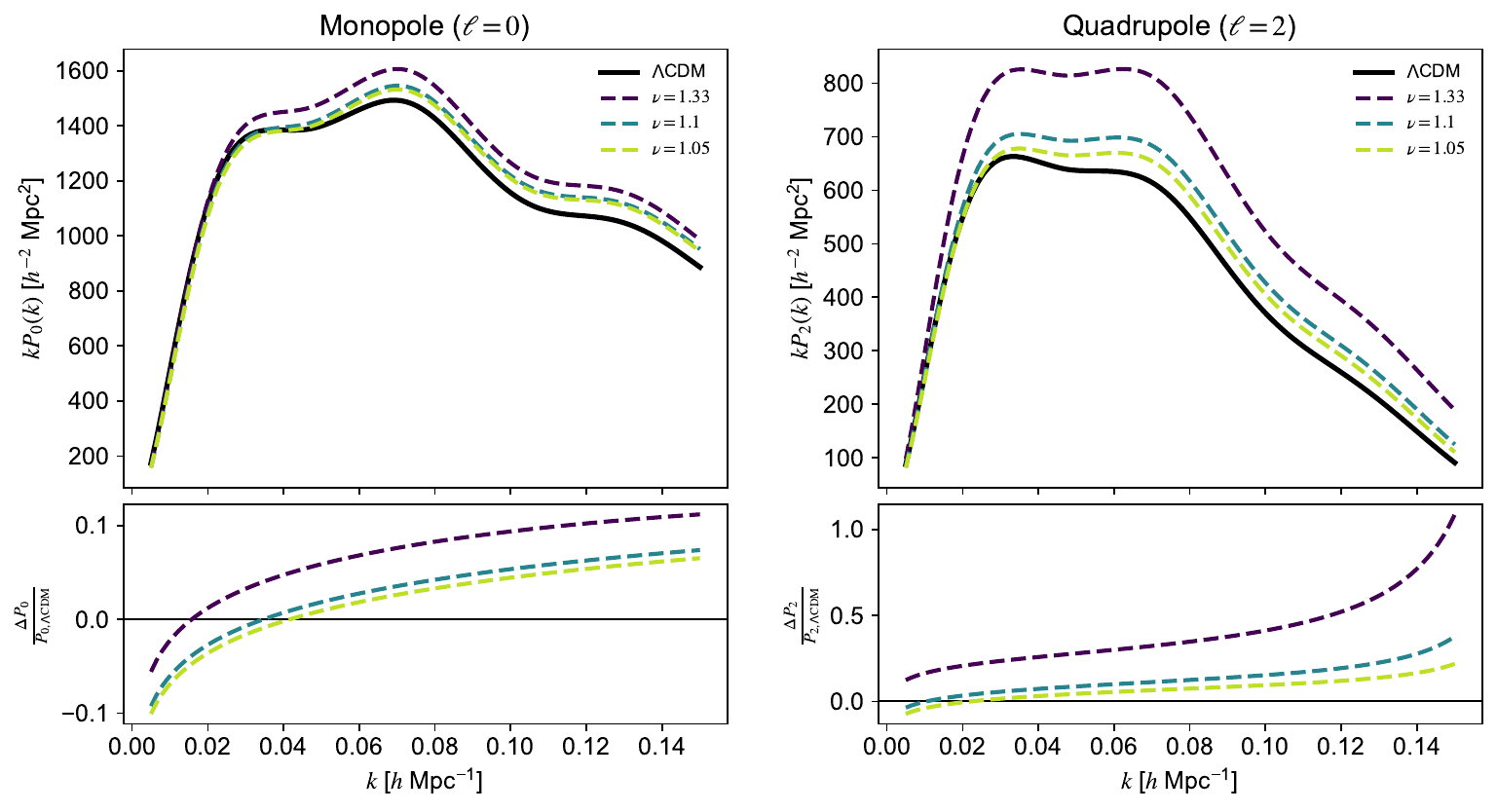}
		\caption{}
		\label{fig_d_4}
	\end{subfigure}
	
	\caption{Power spectra multipoles generated for $n=1.81$. Panel (a) shows for $b=1.8$ and $\sigma_p=3$ Mpc/h, while Panel (b) illustrates for $b=2$ and $\sigma_p=5$ Mpc/h.}
	\label{fig:combined_vertical_2}

\end{figure}

\begin{figure}[!ht]
	\centering
	
	\begin{subfigure}{\textwidth}
		\centering
		\includegraphics[width=\linewidth]{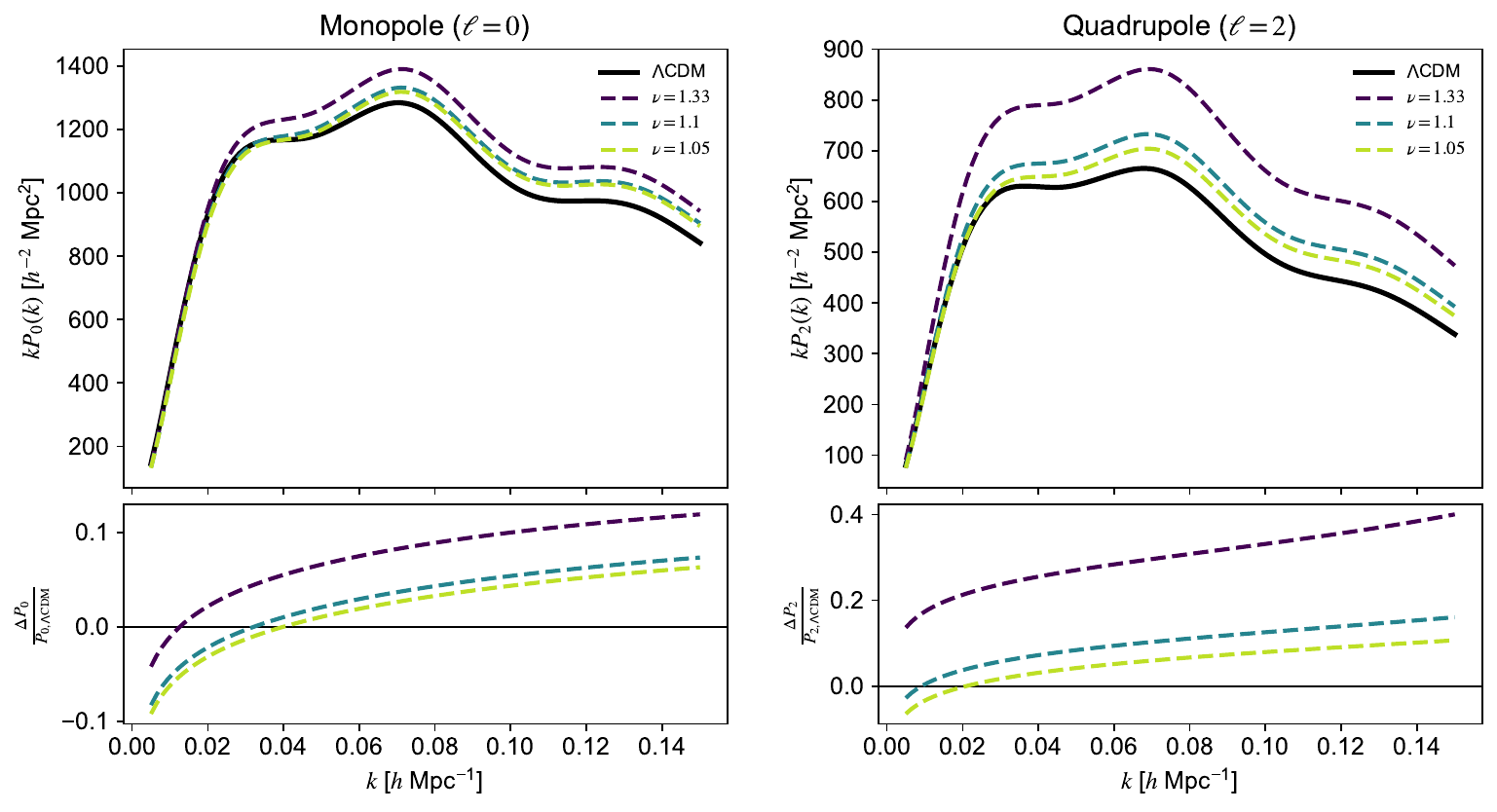}
		\caption{}
		\label{fig_d_5}
	\end{subfigure}
	
	\begin{subfigure}{\textwidth}
		\centering
		\includegraphics[width=\linewidth]{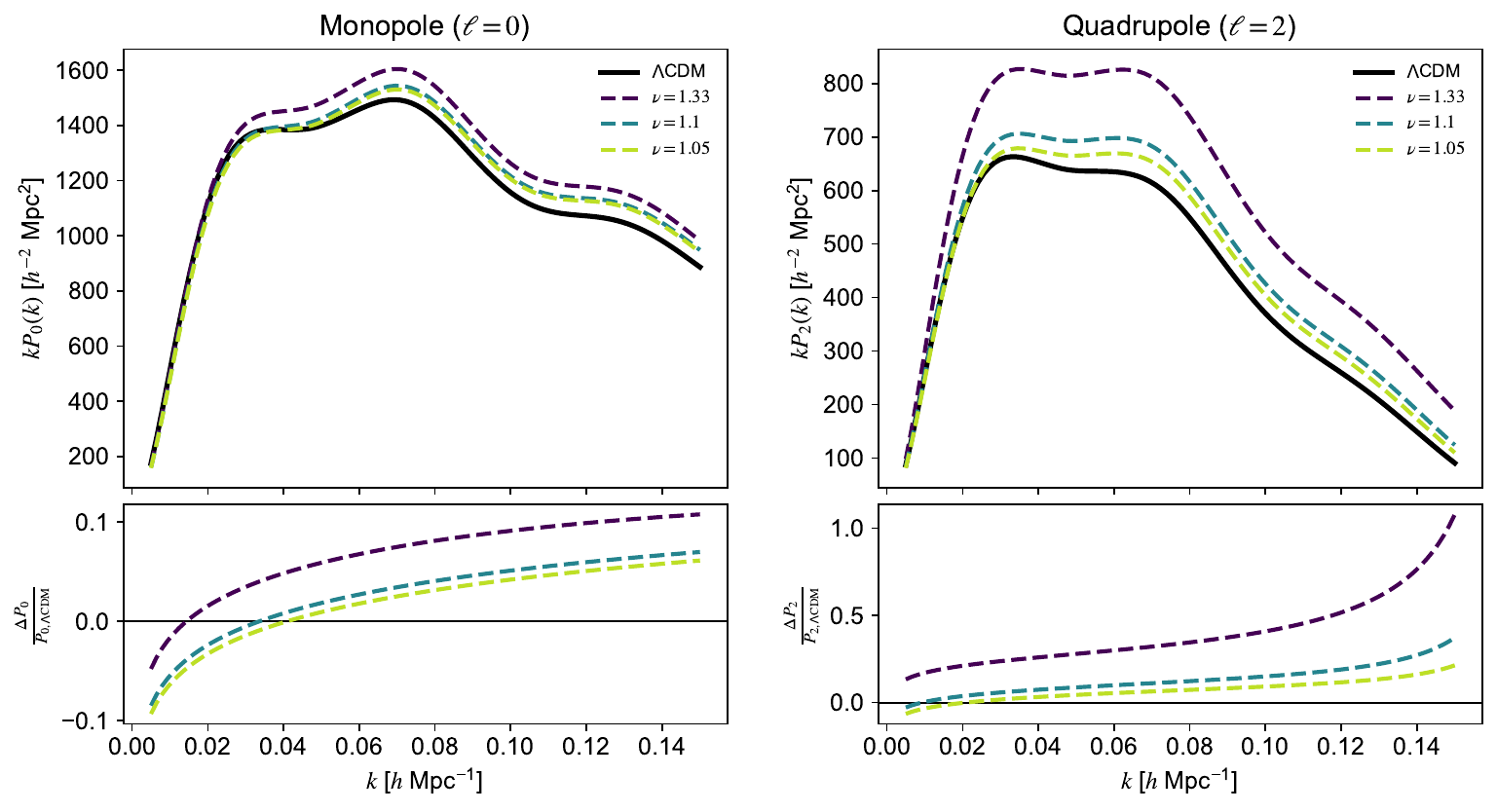}
		\caption{}
		\label{fig_d_6}
	\end{subfigure}
	
	\caption{Power spectra multipoles generated for $n=2.0$. Panel (a) shows for $b=1.8$ and $\sigma_p=3$ Mpc/h, while Panel (b) illustrates for $b=2$ and $\sigma_p=5$ Mpc/h.}
	\label{fig:combined_vertical_3}

\end{figure}

The equation for the redshift-space power spectrum $P_s(k, \beta)$ is given as,
\begin{equation}\label{eq46}
	P_s(k, \beta) = \left( b + f \beta^2 \right)^2 P_\text{lin}(k) D(k, \beta, \sigma_p),
\end{equation}
where, $b$ is the linear galaxy bias (a free parameter). $f$ is the growth rate. In $\Lambda$CDM model it is given by $f \approx \Omega_m^{0.55}$. {In $f(R)$ gravity, this becomes the one expressed in equation~\eqref{mod_f}. It is to be noted that $\nu=1.33 \ (=4/3)$ for $m_\psi=0$ and it gives maximal deviation ($18\%$) of growth rate from $\Lambda$CDM cosmology. We consider this maximal deviation along with some free choices as $\nu=1.10$ ($6\%$ deviation from growth rate) and $\nu=1.05$ ($3\%$ deviation from growth rate) to study matter power spectra at monopole and quadrupole level (see below figures~\ref{fig:combined_vertical_1}--\ref{fig:combined_vertical_3})}. $\beta$ represents the cosine of the angle between the wavevector $k$ and the line-of-sight ($\beta = k_\parallel / k$). $P_\text{lin}(k)$ is the linear matter power spectrum. $\sigma_p=\sigma_v/H_0$ represents pairwise velocity dispersion and $D(k, \beta, \sigma_p)$ is the ``small-scale velocity damping factor".

We consider Gaussian type damping, 
\begin{equation}
	D(k, \beta, \sigma_p) = \exp\left( - (k \beta \sigma_p)^2 \right).
\end{equation}

We cannot measure $P(k, \beta)$ directly, instead we measure the moments. The multipoles $P_\ell(k)$ are projections of $P_s(k, \beta)$ onto Legendre polynomials $\mathcal{L}_\ell(\beta)$,
\begin{equation}
	P_\ell(k) = \frac{2\ell + 1}{2} \int_{-1}^{1} P_s(k, \beta) \mathcal{L}_\ell(\beta) d\beta .
\end{equation}

Using equation~\eqref{eq46}, we can express $P_\ell(k)$ as,
\begin{equation}\label{eq16}
	P_{\ell}(k) = \frac{2\ell + 1}{2} P(k) \int_{-1}^{1} (b + f\beta^2)^2 e^{-(k\beta\sigma_{p})^2} \mathcal{L}_{\ell}(\beta) d\beta.
\end{equation}

To tally with observation we evaluate this for $\ell=0$ (monopole) and $\ell=2$ (quadrupole). For monopole ($\ell=0$): $\mathcal{L}_0(\beta) = 1$ and for quadrupole ($\ell=2$): $\mathcal{L}_2(\beta) = \frac{1}{2}(3\beta^2 - 1)$.

We have calculated the monopoles ($P_0$) and quadrupoles ($P_2$) for both $\Lambda$CDM and $f(R)$ scenario in the $k$ range $10^{-4}-0.15$ (to preserve the linearity). For $\Lambda$CDM, we have generated $P(k)$ using \texttt{CAMB} and the scale independent growth rate $f^{\Lambda\text{CDM}} \approx \Omega^{0.55}$ is considered. For the $f(R)$ gravity scenario, we have implemented the modified power spectra $P_{\delta_{m}}^{f(R)}(k) \propto P_{\delta_{m}}^{\Lambda\text{CDM}}(k) . k^{\frac{\sqrt{33}-5}{6n+4}}$ (see equation~\eqref{eqMPSp}) and modified growth rate as,
\begin{equation}
	f^{f(R)}=f^{\Lambda\text{CDM}}+\frac{\sqrt{1+24 \nu(a,k)}-5}{4}.
\end{equation}

Bias, $b$ and non-linear velocity dispersion, $\sigma_p=\sigma_v/H_0$ are free parameters here and depends on the galaxy samples, however, we have used two sets of values for $b$ and $\sigma_v$ for plotting the $P_0$ and $P_2$ in both $\Lambda$CDM and $f(R)$ gravity scenario. {The theoretical plots showing the monopoles and quadrupoles for different $\nu$ values each for two sets of values of $b=1.8, \ 2$ and $\sigma_v= 300, \ 500 \text{ km s}^{-1}$ for three different values of $n=1,\ 1.81 \text{ and } 2$ are shown in figures~\ref{fig:combined_vertical_1} -- \ref{fig:combined_vertical_3}. The observable shown in these two figures is already available from DESI and more precise measurements will soon be provided by DESI, EUCLID, 4MOST, PFS which has great potential to constrain the parameters of the model}.

The plots also display the fractional deviation of $f(R)$ $P_0$ and $P_2$ from $\Lambda$CDM scenario. It clearly shows that while monopole and quadrupole matter power in $f(R)$ theory increase towards larger $k$ (smaller scales), quadrupole power spectrum remains elevated relative to $\Lambda$CDM for smaller $k$ (larger scales) upto $k\approx0.02$.

\subsection{Transition redshift and scalaron mass}

\begin{figure}[!ht]
	\centering
	\includegraphics[width=0.7\textwidth]{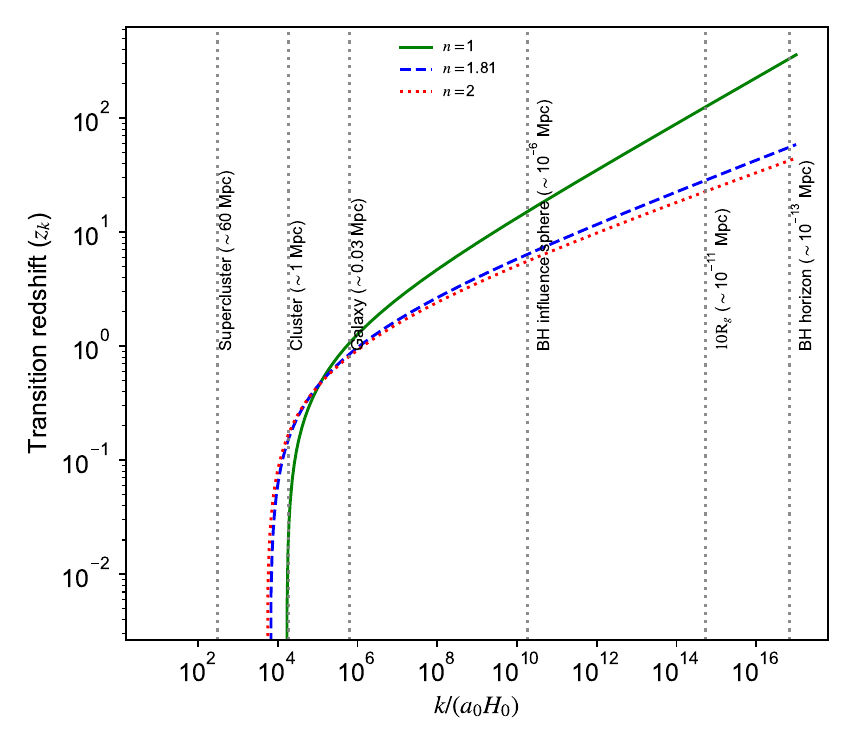}
	\caption{\label{fig2} Transition redshift versus fluctuation mode wavenumber. Colours represent $z_k$ variation for different values of $n$.}
\end{figure}

\begin{figure}
	\centering
	\includegraphics[width=0.7\textwidth]{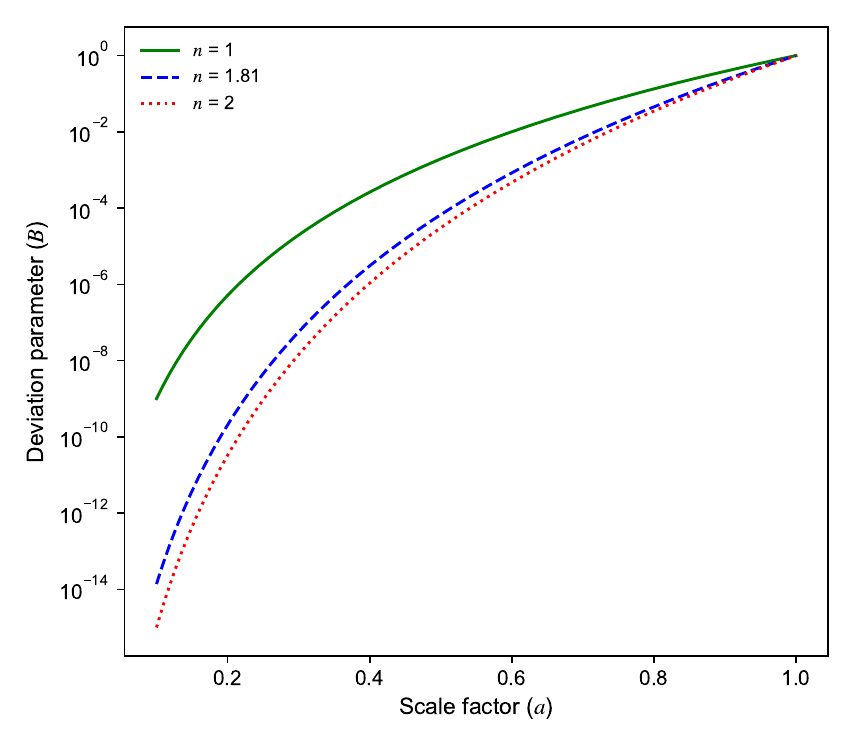}
	\caption{\label{fig3} Variation of deviation parameter ($B$) with scale factor ($a$). Coloured lines show the evolution of the parameter $B$ for different values of $n$.}
\end{figure}

The transition redshifts ($z_k$) of different structures are shown in Figure~\ref{fig2}. The structures span scales from superclusters down to supermassive black holes. We choose scales of massive black holes -- the ones with $M\approx10^6 \text{ M}_\odot$, similar to the mass of Galactic Center black hole. These scales are the event horizon, $R_g=2GM/c^2$, $10 R_g$ and sphere of influence of the black hole ($\sim1$pc). Other scales are galaxies, clusters of galaxies and superclusters. {We see that different structures undergo transition from GR phase ($m_\psi^2>>k^2/a^2$) to non-GR phase ($m_\psi^2<<k^2/a^2$) at different epoch of cosmic histories. The smaller scale structures undergo transition in the early matter dominated era ($z_k>>1$). On the other hand, very large scales spend a considerable amount of time in the GR phase of evolution}. {Structures such as superclusters are transiting to modified gravity phase very close to the present epoch}. The deviation from GR which affects the growth of matter perturbation is quantified by deviation parameter displayed in equation \eqref{eq11}. It varies with cosmic scale factor as, $B\varpropto a^{6n+3}$. It has a scalaron mass dependency as, $B\varpropto m_\psi^{-(6n+3)/(3n+3)}$. A decreasing scalaron mass causes deviation from GR. The growth of deviation with scale factor is shown in Figure~\ref{fig3} for $n=1, \ 1.81 \text{ and }2$. It is seen that modified gravity effect is heavily suppressed in the very early phase ($a<<1$) of the universe. The effective gravitational potential of cosmic structure is measured by the ISW potential displayed in equation \eqref{eq12}. It is evaluated from equations \eqref{eq20} as,
\begin{equation}
	\Phi_-=\frac{3 a^2 H^2}{2k^2}\Omega_m\delta_m.
\end{equation}

\noindent

\begin{figure}[!ht]
	\centering
	\includegraphics[width=0.7\textwidth]{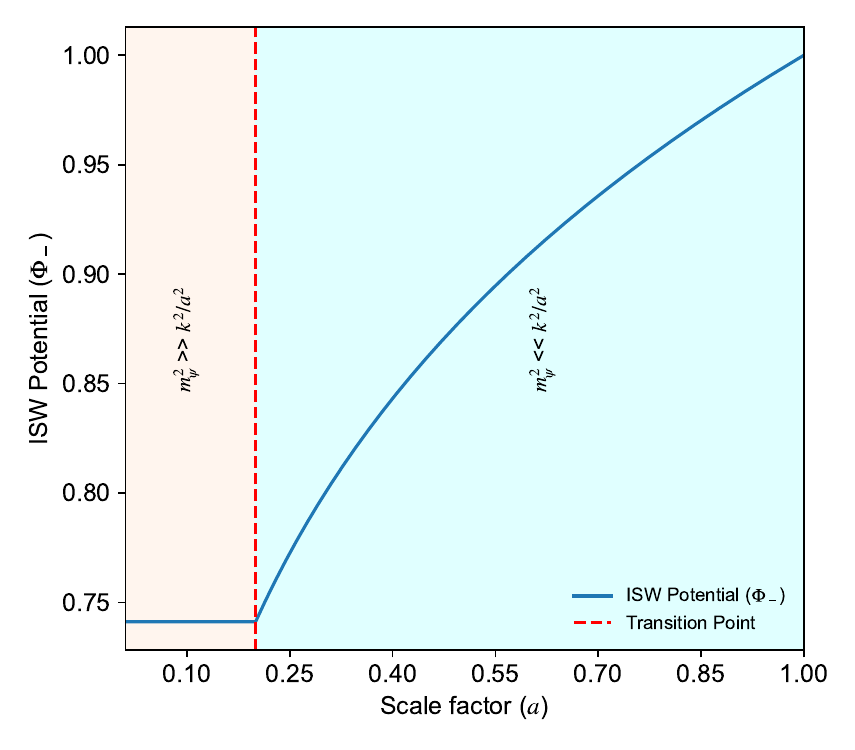}
	\caption{\label{fig4} Variation of ISW potential with cosmic scale factor. The transition epoch represented by the dotted vertical line is estimated from $z_k$ corresponding to fluctuation mode $k\approx 10^{-2} \text { Mpc}^{-1}$.}
\end{figure}

In the GR phase with $H^2 \sim a^{-3}, \ \Omega_m \approx 1 \text{ and } \delta_m \sim a$, the ISW potential remains constant. However, in the modified gravity era, with $\delta_m$ varying according to equation \eqref{eq32}, the ISW potential evolves with cosmic scale factor as, $\Phi_- \varpropto a^{0.186}$. The growth of ISW potential is displayed in Figure~\ref{fig4}. We identify the cause of this growth of potential as decrease of the scalaron mass with cosmic time, as it is evident from the relation $\Phi_- \varpropto m_\psi^{-0.186/(3n+3)}$.

The scalaron masses at transition epoch for various structures are displayed in Table~\ref{table1} for $n=1$ and $n=2$. The transition mass represents the limiting scalaron mass below which evolution of structures proceed via modified gravity channel. We find that transition mass is higher for smaller scale structures. For horizon scales of massive black holes, scalaron mass is in the range $10^{-15}-10^{-14}$ eV (for both the values of $n$). The galactic scale scalaron mass is of the order of $10^{-27}$ eV. We wish to discuss few important aspects of these masses in Section~\ref{sec4}.

\begin{table}
	\centering
	\begin{tabular}{l c c}
		\hline
		\multirow{2}{*}{Cosmic Structures} & \multicolumn{2}{c}{Scalaron masses (eV)} \\
		\cline{2-3}
		& $n=1$ & $n=2$ \\
		\hline
		BH horizon ($R_g$) ($\sim 10^{-13}$ Mpc) & $3.46 \times 10^{-14}$ & $4.52 \times 10^{-15}$ \\
		$10 \, R_g$ ($\sim 10^{-11} $ Mpc) & $1.00 \times 10^{-16}$ & $1.90 \times 10^{-17}$ \\
		BH sphere of influence ($\sim 10^{-6} $ Mpc) & $4.54 \times 10^{-22}$ & $1.84 \times 10^{-22}$ \\
		Galaxy ($\sim 0.03 $ Mpc) & $1.92 \times 10^{-27}$ & $1.69 \times 10^{-27}$ \\
		Cluster ($\sim 1 $ Mpc) & $2.86 \times 10^{-29}$ & $3.27 \times 10^{-29}$ \\
		Supercluster ($\sim 60 $ Mpc) & $2.10 \times 10^{-31}$ & $3.27 \times 10^{-31}$ \\
		\hline
	\end{tabular}
	\caption{A table showing scalaron masses for different cosmic structures. Here MBH is the abbreviation for Massive Black Hole. The corresponding quantities are evaluated for the MBH at the Galactic Center.}
	\label{table1}
\end{table}

Time variation of scalaron mass shown in equation \eqref{eq22} can be rewritten with the following normalization,
\begin{equation}
	m_\psi=m_\psi^{(0)} \left(\frac{t}{t_0}\right)^{-2(n+1)},
\end{equation}
where,  $m_\psi^{(0)}$ and $t_0$ are the present value of scalaron mass and age of the universe respectively. For $f(R)$ gravity models considered in this work, the scalaron mass (see equation \eqref{eq22}) takes the analytical form,
\begin{equation}\label{eq47}
	m_\psi=\left[\frac{R_c}{6n(2n+1)\mu}\left(\frac{R}{R_c}\right)^{2(n+1)}\right]^{1/2}.
\end{equation}

Considering flat $\Lambda$CDM cosmology for $H^2$, we get the present value of Ricci scalar as,
\begin{equation}
	R_0=3H_0^2(4-3\Omega_{m}^{(0)}).
\end{equation}

\begin{figure}[!ht]
	\centering
	\includegraphics[width=0.7\textwidth]{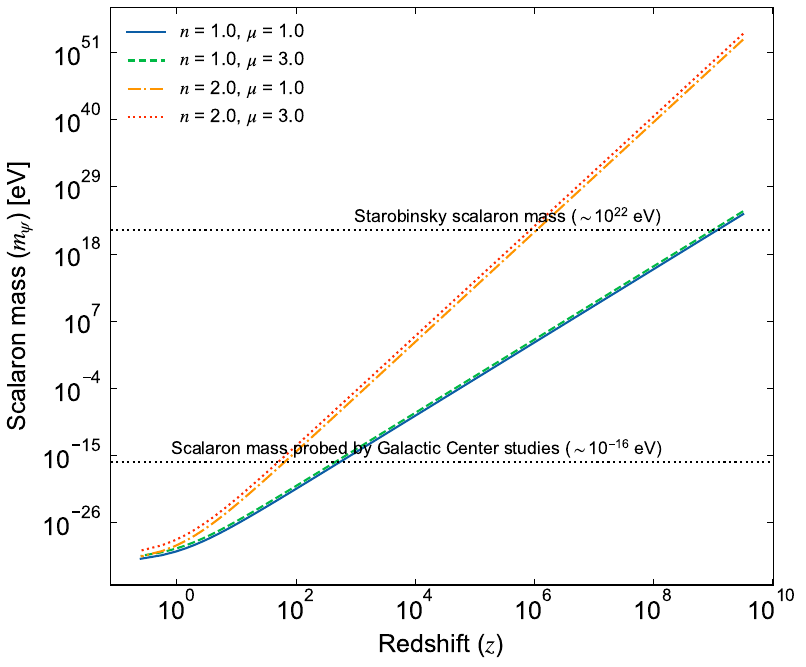}
	\caption{\label{fig5} Evolution of scalaron mass. Coloured lines show evolutions for different model parameters. Black horizontal dotted lines show the two mass scales namely Starobinsky and the scalaron mass that can be probed by the Galactic Centre studies.}
\end{figure}

This gives the present value of scalaron mass as,
\begin{equation}
	m_\psi^{(0)}=\left[\frac{R_c}{6n(2n+1)\mu}\left(\frac{3H_0^2(4-3\Omega_{m}^{(0)})}{R_c}\right)^{2(n+1)}\right]^{1/2},
\end{equation}  
where, $R_c=6H_0^2\Omega_\text{DE}^{(0)}/\mu$.

Figure~\ref{fig5} shows the variation of the scalaron mass with redshift $(z)$ for different choices of model parameters $n$ and $\mu$. The upper dotted horizontal line represents the value of scalaron mass obtained from the Starobinsky inflationary potential \citep{2015JCAP...07..050E} using the observed value of amplitude of scalar power spectrum \citep{2016A&A...594A..13P, 2016A&A...594A..20P}. It is to be noted that the scalaron mass intersects with Starobinsky mass $(\approx 10^{22} \text{ eV})$ at redshift $z \approx 10^{6}$ and $z\approx10^{9}$ for $f(R)$ gravity parameters $n=1$  and $n=2$ respectively. The scalaron mass comes down to $10^{-16}$ eV (see the lower dotted horizontal line) within redshift $z=100-1000$. This mass scale of scalarons has been probed by Galactic Center studies \citep{2020ApJ...893...31K, 2023IJMPD..3250021P, 2024ApJ...964..127P}. We note that the present value of scalaron mass approaches $10^{-33}-10^{-32}$ eV, which is tantalizingly close to the scale of the present Hubble parameter.

\begin{figure}[!ht]
	\centering
	\includegraphics[width=0.7\textwidth]{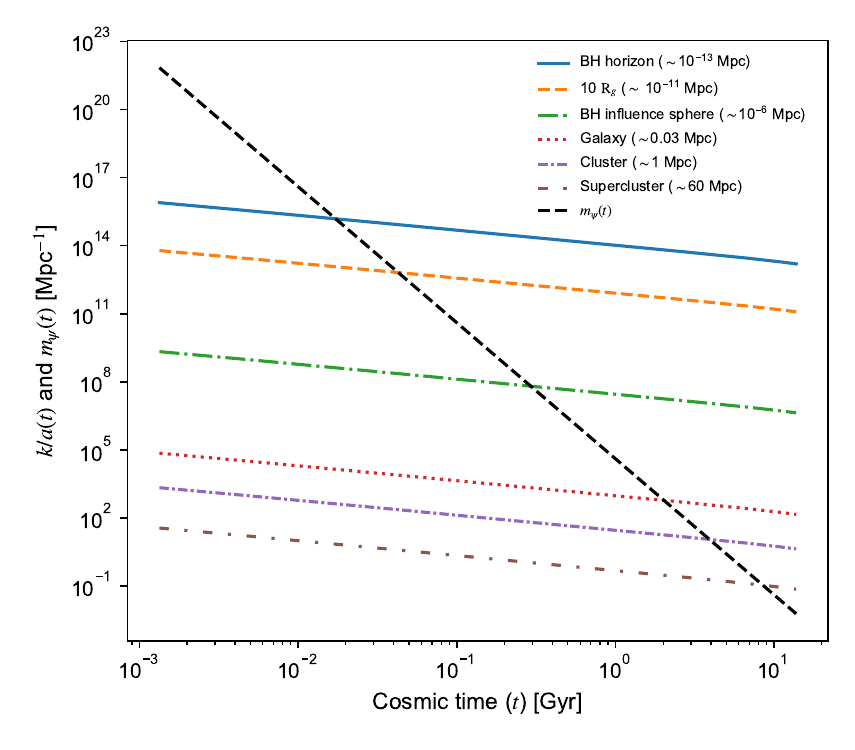}
	\caption{\label{fig6} Variation of $k/a$ and $m_\psi$ with cosmic time.}
\end{figure}

The combined variation of $m_\psi$ and $k/a$ with cosmic time $t$ is shown in figure~\ref{fig6}.

\subsection{Compatibility of evolution of scalaron mass }

As scalaron mass rises towards the past in cosmic history the modified gravity theory asymptotically approaches GR. However, the scalaron mass square must satisfy the constraint ${R}/{m_{\psi }^2}\rightarrow 0$ where $R$ is the background Ricci curvature; scalaron mass must increase faster than the Ricci curvature towards the past \citep{Thomas:2011pj}. This is required to preserve the successful completion of Big Bang Nucleosynthesis (BBN) era which is possible only if ${H^2}/{m_{\psi }^2}\rightarrow 0$ as $t \rightarrow 0$. The range of scalaron fifth force \citep{2018ApJ...855...70K} characterised by the Compton wavelength $m_{\psi }^{-1}$ must shrink faster than the Hubble radius $H^{-1}$ not to spoil the physical processes of the BBN. This is ensured by the time evolution equations $m_{\psi }\propto t^{-2(n+1)}$ and $R\propto H^2\propto t^{-2}$ irrespective of the value of $n$.

\subsection{Scalaron as dark matter and dark energy }

The cosmic evolution of the scalaron mass yields interesting values. The sub galactic scalaron mass of $10^{-22}$ eV realised in the scale of massive black hole sphere of influence resembles the lower end of Fuzzy Dark Matter (FDM) candidates \citep{Widmark:2023dec}. The mass scale $10^{-21} - 10^{-11}$ eV is reserved for axion like dark matter if it has to contribute significantly to dark matter mass budget \citep{Chadha-Day:2021szb}. In this case the scalaron mass scale $10^{-17} -10^{-14}$ eV displayed  by massive black hole horizon scales $(R_g - 10R_g)$ (see Table~\ref{table1}) qualifies as axion like dark matter candidates. The black hole horizon scale scalaron mass of $10^{-16}$ eV is worth mentioning here. In addition to producing general relativistic observables near the Galactic Center black hole (as mentioned earlier) $10^{-16}$ eV scalarons also mimicks axion-like dark matter candidate. All other values of scalaron mass displayed in Table~\ref{table1} are representing minimum bound of scalaron mass in the scales concerned if they evolve via GR like evolutionary paths. Therefore, these scalaron masses are general relativistic limits of scalaron mass in those scales.

In remote past of cosmic history the scalaron mass naturally touches conventional WIMP dark matter mass $\sim 100-1000$ GeV, while at the same time ensuring general relativistic cosmic evolution. As mentioned earlier $10^{-33}$ eV scalarons look like light quintessence which is thought to be responsible for cosmic acceleration. Therefore, as the scalaron mass decreases from remote past to remote future it changes its role from dark matter to dark energy.  We put a remark on this in section~\ref{sec4}.

\section{\label{sec4}Results and Discussions}

In this manuscript we have studied the matter perturbations with scalarons of $f(R)$ gravity theories. For that purpose, we have considered one $f(R)$ gravity Lagrangian which reproduces $\Lambda$CDM like behavior at high redshift (high curvature regime). This predicts a transition from GR like matter era to a non-GR phase of cosmogenic evolution. From scale dependence of the transition epoch, we predict GR limit of scalaron masses for massive black hole scales, galaxies, clusters of galaxies and superclusters. The main results are discussed below.

The non-GR phase of evolution is characterized by enhanced matter perturbation power spectrum. It is found that the power spectrum is enhanced relative to the $\Lambda$CDM prediction for perturbation mode $k>10^{-1} \text{ Mpc}^{-1}$. This is found to be possible for the parameter $n$ of the $f(R)$ gravity Lagrangian constrained by observed departure from Harrison-Zeldovich scale invariant spectrum. The multipole power spectra have been calculated in $f(R)$ gravity theory and compared with the $\Lambda$CDM scenario so as to make large scale structure observables ready for upcoming measurements of DESI, EUCLID, 4MOST and PFS. {We find that $f(R)$ monopole and quadrupole matter power increase towards larger $k$ (smaller scales). Quadrupole matter power spectrum remains elevated relative to $\Lambda$CDM for smaller $k$ (larger scales) upto $k\approx0.02$ (See figures~\ref{fig:combined_vertical_1} --~\ref{fig:combined_vertical_3})}. 

We observe that transition epoch for galaxies and massive black holes lies in deep matter era. Therefore, any growth of these structures in late times must be able to experience modified gravity effects. Evolution of black hole sphere of influence above $z=5-15$ and galaxies above $z=0.8-1$ are strictly general relativistic. {Very large scale structures such as superclusters are transiting to modified gravity phase very close to the present epoch.}

The deviation from GR is caused by a decreasing scalaron mass in cosmic expansion history as is evident from the growth of deviation parameter and ISW potential displayed in Figures~\ref{fig3} and~\ref{fig4}. We calculate the scalaron masses at the transition epoch. As the $f(R)$ Lagrangian is very close to $\Lambda$CDM, the calculated scalaron mass at transition epoch corresponds to GR limit of $f(R)$ theories. It is, therefore, expected that these scalaron masses do not spoil tests of GR carried out or to be carried out in these scales concerned. We put remark on scalaron masses of two scales -- massive black hole horizon and galactic scale. The scalaron mass of $10^{-16}$ eV has been found to reproduce general relativistic periapsis shift of compact orbits of S-stars near the Galactic Center black hole \citep{2020ApJ...893...31K, 2023IJMPD..3250021P}. The impact of scalarons on stationary, axisymmetric vacuum solution of $f(R)$ gravity field equations has shown that for $10^{-16}$ eV scalaron the size of the shadow of the Galactic Center black hole is impressively close to the Kerr black hole prediction \citep{2024ApJ...964..127P} (see the horizontal line representing this mass scale in Figure \ref{fig5}). It ensures this mass scale as appropriate GR limit of $f(R)$ gravity theory in the black hole scale.

The Compton wavelength of $10^{-27}$ eV scalarons fits with galactic size (a few tens of kpc). If galactic scale test of GR is successful, this corresponds to lower bound of galactic scale scalaron mass.

Evolution of scalaron mass (see Figure~\ref{fig5}) shows that the condition $({R}/{m_{\psi }^2})\sim({H^2}/{m_{\psi }^2})\rightarrow 0$ in the early universe for a successful completion of the general relativistic processes such as the BBN is respected by the scalaron mass variation represented by equation~\eqref{eq22} irrespective of the model parameter $n$. Scalaron mass touches the Starobinsky mass ($\approx10^{22}$ eV) at redshifts $z\approx10^6$ and $z\approx10^{9}$ respectively for model parameters $n=1 \text{ and } 2$. At present time ($z\approx 0$), scalaron mass comes down to $10^{-33}-10^{-32}$ eV. This occurs for different choices of $n$ and $\mu$ appearing in the $f(R)$ gravity Lagrangian. This mass scale is tantalizingly close to the scale of the Hubble parameter, $H_0$. In a purely de-Sitter phase of late time expansion history of the universe, this mass scale is naturally very close to the cosmological constant, $\Lambda \sim m_{\psi(0)}^2$. Therefore, present scalaron mass qualifies as a candidate of light quintessence responsible for accelerated expansion. We reproduce a light scalar mass through cosmological evolution, rather than by mere adjustment of a scalar field theory to obtain the criterion for late time cosmic acceleration. Equation \eqref{eq47} implies that $m_\psi \varpropto \rho^{n+1}$. This indicates that high redshift observations in cosmology including the ones involving CMB $(z\sim 3000)$ are not likely to give clues of non-GR like cosmogenic evolution. Any deviation from GR has to be investigated through late time clustering of structures such as massive black holes, galaxies and other large-scale structures.

We end with few remarks on scalaron masses displayed in Table~\ref{table1}. These scalaron masses are general relativistic limits of $f(R)$ gravity theory. At the same time scalarons with mass range $10^{- 22}  - 10^{-14}$ eV realised in the scales of black hole sphere of influence and horizon scales resemble interesting dark matter particle candidates such as FDM and axion like particles. WIMP mass scale is natural consequence of the mass variation model of scalarons. Therefore, these scalaron masses serve both as dark matter candidates and general relativistic limit of a modified gravity theory. Realisation of dark matter like effect in a theory beyond GR is possible. This is unlike a school of thought which advocates for replacing dark matter particle altogether in modified gravity theory. In addition, realisation of light quintessence mass $m_{\psi }\sim\sqrt{\Lambda }\sim H_0$ in the present epoch inspires one to treat scalaron as a dark energy candidate.

\bibliography{refs}{}
\bibliographystyle{unsrtnat}
\end{document}